\documentclass[prd,showpacs]{revtex4}

\usepackage{bm}
\usepackage{amsmath}
\usepackage{natbib}

\def\ba{{\bm a}}
\def\bb{{\bm b}}
\def\bc{{\bm c}}

\def\bk{{\bm k}}

\def\bl{{\bm l}}

\def\bn{{\bm n}}
\def\bN{{\bm N}}

\def\bp{{\bm p}}
\def\bq{{\bm q}}

\def\bx{{\bm x}}

\def\l{\lambda}
\def\bla{{\bm \lambda}}

\def\bna{{\bm \nabla}}

\def\bDa{{\bm \Delta}}

\def\rab{\mbox{ $R_{AB}$}}

\def\lb{\label}

\def\be{\begin{equation}}
\def\ee{\end{equation}}
\def\bea{\begin{eqnarray}}
\def\eea{\end{eqnarray}}

\begin{document}

\title{Influence of mass multipole moments on the deflection of a light ray \\
by an isolated axisymmetric body}

\author{Christophe Le Poncin-Lafitte}
\email{christophe.leponcin-lafitte@obspm.fr}
\affiliation{Lohrmann Observatory, Dresden Technical University,\\ Mommsenstr. 13, D-01062 Dresden, Germany}

\author{Pierre Teyssandier}
\email{Pierre.Teyssandier@obspm.fr}
\affiliation{D\'epartement Syst\`emes de R\'ef\'erence Temps et Espace,
CNRS/UMR 8630, \\
Observatoire de Paris, 61 avenue de l'Observatoire, F-75014 Paris, France}

\date{\today}

\begin{abstract}
Future space astrometry missions are planned to measure positions and/or parallaxes of celestial objects with an accuracy of the order of the microarcsecond. At such a level of accuracy, it will be indispensable to take into account the influence of the mass multipole structure of the giant planets on the bending of light rays. Within the parametrized post-Newtonian formalism, we present an algorithmic procedure enabling to determine explicitly this influence on a light ray connecting two points located at a finite distance. Then we specialize our formulae in the cases where 1) the light source is located at space infinity, 2) both the light source and the observer are located at space infinity. We examine in detail the cases where the unperturbed ray is in the equatorial plane or in a meridian plane.
\end{abstract}

\pacs{04.20.Cv, 04.25.Nx, 04.80.-y, 95.55.Br}

\maketitle

\section{Introduction}

\par Highly precise astrometry and tests of general relativity in the Solar System will require in the foreseeable future the measurements of apparent positions of light sources with an accuracy of the order of one microarcsecond ($\mu$as) or better. The Global Astrometric Interferometer for Astrophysics (GAIA) Mission is planned to obtain accuracies in the range 4-20 $\mu$as \cite{GAIA:2000,Bienayme:Turon:2002} and the Space Interferometer Mission (SIM) is designed to reach a differential accuracy of $0.6$ $\mu$as on bright stars \cite{unwin}.  

\par This context has stimulated several studies devoted to the bending effects due to the mass multipole moments of the bodies of the Solar System \cite{Hellings,Klioner:1992,Klioner:2003a,Angonin,2006CQGra..23.4853C,Kopeikin:1997,2006CQGra..23.4299K,2007PhRvD..75f2002K}. Owing to the complexity of the calculations, only the contribution due to the quadrupole moment $J_{2}$ of the deflecting mass has been explicitly determined. For a light ray grazing Jupiter, it has been shown that this effect amounts to 240 $\mu$as \cite{Klioner:2003a} and may be proposed as a new test of general relativity which could be realized with GAIA \cite{2006CQGra..23.4853C,2007PhRvD..75f2002K}. However, rough estimates show that the Jovian higher multipole moments $J_{4}$ and $J_{6}$ could produce deflections of the order of $10$ $\mu$as and $0.1$ $\mu$as, respectively. For this reason, we propose here a method enabling to determine the deflection effects of any mass multipole moment of a static axisymmetric gravitational field within the parametrized post-Newtonian formalism. This method is based on an algorithmic procedure found in \cite{Kopeikin:1997} and \cite{linet1}. The problem is treated in the general case where both the light source and the observer are located at a finite distance. It is worthy of note that being a development of results derived in \cite{linet1}, \cite{leponcin:2004} and \cite{teyssandier} by using the so-called Synge's world function, our procedure avoids any integration of the null geodesic equations. 

\par The paper is organized as follows. In Sec. II we give the notations used in this article. In Sec. III we recall how in a static axisymmetric space-time the vector tangent to a null geodesic can be derived from the time transfer function ${\cal T}$ giving the travel time of a photon between two points. In Sec. IV we determine the contribution ${\mathcal T}_{J_{n}}$ of each mass multipole $J_{n}$ to the time transfer function ${\cal T}$. In Sec. V we carry out the calculation of the contribution of each mass multipole to the direction of a light ray at its point of reception, both the light source and the observer being assumed to be located at a finite distance. In Sec. VI we focus on the case where the source of the light ray is located at space infinity. Indeed, this assumption is sufficient for the light emitted by stars or by extragalactic objects. In Sec. VII we assume that both the source of light and the observer are situated at space infinity. We obtain explicit formulae for the contribution of each mass multipole to the light deflection. We carry out a detailed calculation of the influence of $J_{2}$, $J_{3}$ and $J_{4}$. We recover the results on the effects of $J_{2}$ previously obtained in \cite{Klioner:1992,Klioner:2003a,Kopeikin:1997,2006CQGra..23.4299K,2007PhRvD..75f2002K} and \cite{2006CQGra..23.4853C}. We give some concluding remarks in Sec. VIII.

\section{Notations and conventions}

\par The Lorentzian metric of space-time is denoted by $g$. The signature adopted for $g$ is $(+\;-\;-\;-)$. We suppose that space-time is covered by a global quasi-Galilean coordinate system $(x^\mu)=(x^0,{\bx})$, where $x^0=ct$, $t$ being a time coordinate, and ${\bx}=(x^i)$. Greek indices run from 0 to 3, and latin indices run from 1 to 3. 

A bold letter denotes an ordered triple. In order to distinguish the triples built with contravariant components of a vector from the ones built with covariant components, we systematically use the notation ${\ba} =(a^1,a^2,a^3)=(a^i)$ and $\underline{\bb}= (b_1,b_2,b_3)=(b_i)$, except for the gradient operator, denoted by $\bna$ as usual. Let us emphasize that these notations are purely conventional and that we shall write for example $\underline{\bl} = \bN + \underline{\bla}$ in order to mean that $l_{i} = N^{i} + \lambda_{i}$. Such a mixing of ``contravariant" and ``covariant" quantities cannot be ambiguous, since these equalities must be understood as equalities of functions, and not as equalities between true four dimensional vectors. Given $\ba = (a^i), \bb = (b^i)$ and $\underline{\bc}  = (c_i)$, we use $\ba.\bb$ to denote $a^i b^i$ and $\ba.\underline{\bc}$ to denote $a^i c_i$ (Einstein's convention on repeated indices is used in both cases). We put $\vert \ba \vert = (\delta_{ij}a^{i}a^{j})^{1/2}$ and $\vert \underline{\bb}\vert = (\delta^{ij}b_{i}b_{j})^{1/2}$. 

$G$ is the Newtonian gravitational constant and $c$ is the speed of light in a vacuum. 

\section{Time transfer function and direction of a light ray}

We assume that the metric does not depend on $x^0$. Let $x_A=(ct_A,\bx_A)$ and $x_B=(ct_B,\bx_B)$ be two events of space-time supposed to be connected by an unique light ray $\Gamma_{AB}$. By convention $x_A$ and $x_B$ denote the emission point and the reception point, respectively. The travel time $t_B-t_A$ of a photon connecting $x_A$ and $x_B$ is a function of $\bx_A$ and $\bx_B$, so we put
\be \lb{4}
t_B - t_A = {\cal T} (\bx_{A}, \bx_{B}).
\ee

We call ${\cal T}$ the time transfer function (relative to the chosen coordinate system). Following a theorem shown in \cite{leponcin:2004}, the direction tangent to the light ray $\Gamma_{AB}$ at point $\bx_B$ is defined by the 4-vector having the covariant components
\be \lb{5}
(l_0)_B = 1, \quad (l_i)_B =
- \, c\, \frac{\partial {\cal T}}{\partial x_{B}^{i}}(\bx_A,\bx_B). 
\ee

We suppose that the gravitational field is generated by an isolated axisymmetric body. We are only interested in calculating the contributions of the mass multipoles to the bending of light at the order $1/c^2$. So using a standard post-Newtonian gauge \cite{Will:1993} we may content ourselves with a metric given by
\bea
g_{00} &=& 1 - \frac{2W}{c^2} +  {\cal O}\left(\frac{1}{c^4}\right), \\
g_{0i} &=& {\cal O}\left(\frac{1}{c^3}\right), \\
g_{ij} &=& -\left(1 + \gamma \, \frac{2W}{c^2}\right)\delta_{ij} + {\cal O}\left(\frac{1}{c^4}\right),  
\eea
where $W$ is a potential defined by 
\be \lb{pot}
W(\bx) = G\int \frac{\rho(\bx')}{\vert \bx - \bx' \vert }d^{3}\bx' + {\cal O}\left(\frac{1}{c^2}\right),
\ee
$\rho$ being the rest mass density of the body and $\gamma$ the well-known post-Newtonian parameter which describes the three-space curvature. Putting   \be \lb{RN}
R_{AB} = \mid \! \bx_B- \bx_A \! \mid \, , \qquad \bN_{AB} = \frac{\bx_B - \bx_A}{R_{AB}},
\ee
the time transfer function ${\mathcal T}(\bx_{A}, \bx_{B})$ involved in Eq. (\ref{5}) is then determined by (see, e.g., Ref. \cite{linet1})
\be \lb{6n}
{\mathcal T}(\bx_{A}, \bx_{B})  =  \frac{1}{c}R_{AB} + \frac{1}{c^3}(\gamma +1)R_{AB}\int_{0}^{1}W(\bx_{(0)}(\lambda)) d\lambda 
+ {\cal O}\left(\frac{1}{c^4}\right),
\ee
the integral being calculated along the segment of a line of ends $\bx_A$ and $\bx_B$ described by the parametric equations
\be \lb{9}
\bx_{(0)}(\lambda) = R_{AB} \bN_{AB} \lambda + \bx_{A} \, , \qquad 0 \leq \l \leq 1.
\ee

\section{Multipole structure of the time transfer function}

Throughout this work, the center of mass $O$ of the body is taken as the origin of the quasi-Cartesian coordinates $x^{i}$ and the axis of symmetry is chosen as the $x^3$-axis. We put
\be \lb{r}
r=\mid \! \bx \! \mid \, , \quad r_A=\mid \! \bx_A \! \mid \, , \quad r_B=\mid \! \bx_B \! \mid.
\ee

We assume that the smallest sphere centered on O and containing the body has a radius equal to the equatorial radius $r_e$ of the body and that the segment joining $\bx_A$ and $\bx_B$ is outside this sphere. At any point $\bx$ such that $r \geq r_{e}$ $W(\bx)$ is then given by the multipole expansion
\be \lb{W}
W(\bx) = \frac{GM}{r}\left[1 - \sum_{n=2}^{\infty} J_n \left(\frac{r_e}{r}\right)^n P_n \left(\frac{\bk .\bx}{r}\right)\right],
\ee
where $\bk$ denotes the unit vector along the $x^3$-axis, the $P_{n}$ are the Legendre polynomials, $M$ is the mass of the body and the coefficients $J_n$ are the mass multipole moments. As a consequence the integral involved in the right-hand side (r.h.s.) of Eq. (\ref{6n}) may be written as (see Refs. \cite{Kopeikin:1997} and \cite{linet1})
\be 
\lb{39a}
\int_{0}^{1}W\left( \bx_{(0)}(\lambda )\right) d\lambda =
GM\left[ 1-\sum_{n=2}^{\infty}\frac{1}{n!}J_nr_{e}^{n}\frac{\partial^n}{(\partial x^3)^n}\right]F(\bx ,\bx_A,\bx_B)\bigg|_{\bx =0},
\ee
where $F(\bx,\bx_A,\bx_B)$ is the Shapiro kernel function defined by
\be \lb{F}
F(\bx ,\bx_A,\bx_B)=\frac{1}{R_{AB}}\ln \left( \frac{\mid \! \bx -\bx_A\! \mid
+\mid \! \bx -\bx_B\! \mid+R_{AB}}{\mid \! \bx -\bx_A\! \mid +
\mid \! \bx -\bx_B\! \mid -R_{AB}} \right).
\ee

Substituting Eq. (\ref{39a}) into Eq. (\ref{6n}), and then using Eq. (\ref{F}) yield the expansion  
\be 
\label{TT}
\mathcal{T}(\bx_A,\bx_B)=\frac{1}{c}\rab+\mathcal{T}_{M}(\bx_A,\bx_B)+\sum_{n=2}^{\infty} \mathcal{T}_{J_n}(\bx_A,\bx_B),
\ee
where ${\mathcal T}_M$ is the well-known Shapiro time delay
\be \lb{M0}
\mathcal{T}_{M}(\bx_A,\bx_B)=(\gamma+1)\frac{GM}{c^3} \ln\left(\frac{r_A+r_B+R_{AB}}{r_A+r_B-R_{AB}}\right)
\ee
and each ${\mathcal{T}_{J_n}}$ is determined by
\be \label{masse}
\mathcal{T}_{J_n}(\bx_A,\bx_B)=-(\gamma+1)\frac{GM}{c^3}\frac{1}{n!}J_n r_e^n \frac{\partial^{n}}{(\partial x^{3})^n}\ln \left( \frac{\mid \! \bx -\bx_A\! \mid
+\mid \! \bx -\bx_B\! \mid+R_{AB}}{\mid \! \bx -\bx_A\! \mid +
\mid \! \bx -\bx_B\! \mid -R_{AB}} \right)\bigg|_{\bx =0}.
\ee

To carry out an explicit calculation of the r.h.s. of Eq. (\ref{masse}), let us apply Fa\`a di Bruno's formula \cite{gradshteyn} giving the $n$th derivative of a composite function $f(x) = h [u(x)]$, namely \footnote{It is easily checked that Eq. (\ref{faa}) is equivalent to the usual form of Fa\`a di Bruno's formula.}
\be \lb{faa}
\frac{d^n}{dx^n}f(x) = \sum_{m=1}^{n} \frac{d^{n-m+1}h(u)}{du^{n-m+1}}  \sideset{}{'}\sum_{i_1,...,i_m} \frac{n!}{i_1!i_2!...i_{m}!} 
\left(\frac{1}{1!}\,\frac{d^{}u}{dx^{}}\right)^{i_{1}} \left(\frac{1}{2!}\,\frac{d^{2}u}{dx^{2}}\right)^{i_{2}} \ldots \left(\frac{1}{m!}
\,\frac{d^{m}u}{dx^{m}}\right)^{i_{m}},
\ee
where $\sideset{}{'}\sum_{i_1,...,i_m}$ denotes the summation over the sets of non negative integers $i_1$, $i_2$, ..., $i_{m}$ satisfying the pair of equations
\bea 
\left\{ \begin{array}{ll}
i_1 + 2 i_2 + 3 i_3 + ... + m i_{m} = n, \lb{bruno23} \\ 
i_1+i_2+...+i_{m} = n - m +1,   \\ 
\end{array} \right.
\eea
with $1 \leq m \leq n$. This calculation involves the $l$th derivatives of $\vert \bx-\bx_A \vert$ and $\vert \bx-\bx_B \vert$ with respect to $x^3$ at $\bx = 0$. It may be seen that
\be \lb{gegen3a}
\frac{1}{l!}\frac{\partial^{\,l} \vert \bx-\bx_A \vert}{(\partial x^{3})^{l}}\bigg\vert_{\bx = 0} = \frac{1}{r_{A}^{l-1}} C_{l}^{(-1/2)}
\left( \frac{\bk.\bx_{A}}{r_{A}}\right) 
\ee
by comparing the Taylor expansion of $\vert \bx-\bx_A  - \delta \bx \vert$ about $\bx$ at point $\bx=0$ with the expansion
\be \lb{gegen2}
\vert \bx-\bx_A  - \delta \bx \vert = \vert \bx-\bx_A \vert \sum_{l=0}^{\infty}\, \left(\frac{\bk . \delta \bx}{\vert \bx-\bx_A \vert}\right)^{l} C_{l}^{(-1/2)}\left( \frac{\bk.(\bx-\bx_A)}{\vert \bx-\bx_A \vert}\right),
\ee
where $\delta \bx = \delta x^{3}\bk$ and $C_{l}^{(-1/2)}$ denotes the Gegenbauer polynomial of degree $l$ and of parameter $-1/2$ (see, e.g., Ref. \cite{abra}). Using Eq. (\ref{gegen3a}) and the similar expression which would be obtained for the $l$th derivatives of $\vert \bx-\bx_B \vert$, Fa\`a di Bruno's formula leads to 
\be \lb{TJn}
{\mathcal T}_{J_n}(\bx_A,\bx_B) = (\gamma+1)\frac{GM}{c^3}\,J_n\, r_e^n \sum_{m=1}^n \left[\frac{1}{(r_A+r_B-R_{AB})^{n-m+1}}-\frac{1}{(r_A+r_B+R_{AB})^{n-m+1}}\right] \Theta_{nm}(\bx_A, \bx_B),
\ee
where $\Theta_{nm}(\bx_A, \bx_B)$ is defined by
\be \lb{Phi}
\Theta_{nm}(\bx_A, \bx_B) = (-1)^{n-m} \sideset{}{'}\sum_{i_1,...,i_m} \frac{(n-m)!}{i_1!i_2!...i_{m}!} \prod_{l=1}^{m}\left[S_{l}(\bx_A, \bx_B)\right]^{i_{l}},
\ee
with
\be \lb{Lk}
S_{l}(\bx_A, \bx_B) = \frac{1}{r_{A}^{l-1}} C_{l}^{(-1/2)}\left(\frac{\bk.\bx_A}{r_A}\right) +
\frac{1}{r_{B}^{l-1}}C_{l}^{(-1/2)}\left( \frac{\bk.\bx_B}{r_B}\right).
\ee

An explicit calculation of each ${\mathcal T}_{J_n}$ is easy. Consider, e.g., the case where $n=2$. The only sets of non negative integers solutions to Eqs. (\ref{bruno23}) are $\{i_1 = 2\}$ for $m=1$ and $\{i_1 =0, i_2 =1\}$ for $m=2$. Then Eqs. (\ref{TJn})-(\ref{Lk}) give
\be \lb{TJ2}
{\mathcal T}_{J_2}(\bx_A,\bx_B) = \frac{\gamma+1}{2}\frac{GM}{c^3}\,\frac{J_2\, r_e^2}{r_A r_B}\frac{R_{AB}}{1 + \bn_A .\bn_B}\left[
\frac{1-(\bk .\bn_A)^2}{r_A} + \frac{1-(\bk .\bn_B)^2}{r_B} - \left(\frac{1}{r_A}+ \frac{1}{r_B}\right)\frac{[\bk . (\bn_A + \bn_B)]^2}{1+\bn_A . \bn_B} \right].
\ee

We thus recover by a straightforward calculation a formula that we have previously derived from the multipole expansion of Synge's world function (see Refs. \cite{linet1} and \cite{teyssandier}).

\section{Direction of a light ray at the reception point}
\par For the sake of brevity, let us use the notation $\underline{\bl}_B=\lbrace(l_i)_B\rbrace$ and put
\be \lb{nAB}
\bn_{A} = \frac{\bx_{A}}{r_{A}}, \qquad \bn_{B} = \frac{\bx_{B}}{r_{B}}.
\ee

Taking into account Eqs. (\ref{TT}), (\ref{M0}) and (\ref{TJn}, Eq. (\ref{5}) yields 
\be \lb{tan} 
\underline{\bl}_B = -\bN_{AB} + \underline{\bla} (\bx_A, \bx_B), 
\ee
where $\underline{\bla} (\bx_A, \bx_B)$ is given by the multipole expansion     
\be
\label{lam}
\underline{\bla} (\bx_A, \bx_B) = \underline{\bla}_{M}(\bx_A,\bx_B) + \sum_{n=2}^\infty \underline{\bla}_{J_n}(\bx_A,\bx_B),
\ee
with
\be \lb{M1}
\underline{\bla}_{M}(\bx_A,\bx_B) =  (\gamma+1)\,\frac{GM}{c^2 r_{B}}\, \frac{1}{1 + \bn_{A} . \bn_{B}}\left[\frac{R_{AB}}{r_{A}}\bn_B -\left(1 + \frac{r_{B}}{r_{A}}\right)\bN_{AB}\right] 
\ee
and
\begin{eqnarray} \lb{laJn}
&&\underline{\bla}_{J_n}(\bx_A,\bx_B) = (\gamma+1)\frac{GM}{c^2}\,J_n r_e^n \sum_{m=1}^{n} \left\lbrace(n - m + 1) \left[\frac{\bn_B - \bN_{AB}}{(r_A+r_B-R_{AB})^{n-m+2}}\right. \right. \nonumber \\
& & \left. \qquad \qquad \qquad \, \, \, \, -\frac{\bn_B + \bN_{AB}}{(r_A+r_B+R_{AB})^{n-m+2}}\right] \Theta_{nm}(\bx_A, \bx_B) \nonumber \\
& & \left. \qquad \qquad \qquad \, \, \, \, + \left[\frac{1}{(r_A+r_B-R_{AB})^{n-m+1}}-\frac{1}{(r_A+r_B+R_{AB})^{n-m+1}}\right]{\bm \Upsilon}_{nm}(\bx_A, \bx_B \right\rbrace, 
\end{eqnarray}
where
\bea \lb{Psi}
& &{\bm \Upsilon}_{nm}(\bx_A, \bx_B) = (-1)^{n-m} \sideset{}{'}\sum_{i_1,...,i_m} \frac{(n-m)!}{i_1!i_2!...i_{m}!} \sum_{l=1}^{m} i_{l} 
\left[S_{l}(\bx_A, \bx_B)\right]^{i_{l}-1}\prod_{\stackrel{q=1,}{q\neq l}}^{m}\left[S_{q}(\bx_A, \bx_B)\right]^{i_{q}} \nonumber \\
& &\qquad \qquad \qquad \quad \, \, \times \frac{\left[ P_{l-1}\left(\bk.\bn_B\right)\bk - P_{l}\left(\bk.\bn_B\right)\bn_B \right]}{r_{B}^{l}}.
\eea 

By convention, $\prod_{\stackrel{q=1,}{q\neq l}}^{m}\left[S_{q}(\bx_A, \bx_B)\right]^{i_{q}} = 1$ when $m=1$.

\par Using the solutions to Eqs. (\ref{bruno23}) found at the end of Sec. IV in the case where $n=2$ it may be seen that Eqs. (\ref{laJn}) and (\ref{Psi}) lead to 
\bea \lb{lJ2}
& &\underline{\bla}_{J_2}(\bx_A,\bx_B) = (\gamma +1)\frac{GM}{c^2} J_2 r_{e}^2 \Bigg\lbrace -[\bk.(\bn_A + \bn_B)]^2 \left[\frac{\bn_B - \bN_{AB}}{(r_A+r_B-R_{AB})^{3}}- \frac{\bn_B + \bN_{AB}}{(r_A+r_B+R_{AB})^{3}}\right]  \nonumber \\ 
& & \quad + \frac{1}{2}\left[\frac{1-(\bk .\bn_A)^2}{r_A} + \frac{1-(\bk .\bn_B)^2}{r_B}\right]\left[\frac{\bn_B - \bN_{AB}}{(r_A+r_B-R_{AB})^{2}} - \frac{\bn_B + \bN_{AB}}{(r_A+r_B+R_{AB})^{2}}\right] \nonumber \\
& & \quad + \frac{1}{r_B^3}\frac{(r_A + r_B)R_{AB}}{r_{A}^{2}}\frac{\bk . (\bn_A + \bn_B)}{(1 + \bn_A . \bn_B)^2}\left[\bk - (\bk . \bn_B)\bn_B\right] +\frac{1}{2 r_B^3}\frac{R_{AB}}{r_A}\frac{2(\bk . \bn_B)\bk + \left[1 - 3(\bk . \bn_B)^2\right]\bn_B}{1 + \bn_A . \bn_B}\Bigg\rbrace.
\eea

A tedious but straightforward calculation shows that this formula is equivalent to the r.h.s. of Eq. (100) in Ref. \cite{teyssandier}. 

\section{Source located at infinity}

\par Given a unit vector $\bN$, let ${\cal D}$ be the straight line parallel to $\bN$ passing through $\bx_{B}$. Suppose that point $\bx_{A}$ is moved away from $\bx_{B}$ along ${\cal D}$ so that $\lim_{r_A \rightarrow \infty}\bn_A = \bN$. This limit corresponds to a source at infinity observed at $\bx_{B}$ in the direction $\bN$ within the zeroth-order approximation. Then $\underline{\bla}(\bx_A, \bx_B)$ becomes a function of $\bN$ and $\bx_{B}$. So we put
\be \lb{lNB}
\underline{\bla} (\bN , \bx_B) = \lim_{r_A \rightarrow \infty, \, \bn_A \rightarrow \bN} \underline{\bla}(\bx_{A}, \bx_B).
\ee 

Let 
\be \lb{csa}
\cos \alpha = -\bn_{B}. \bN , \qquad 0\leq \alpha <\pi
\ee
and denote by $r_{c}$ the quantity
\be\label{imp}
r_{c} = r_{B} \sin \alpha ,
\ee
that is the impact parameter of the light ray at the zeroth-order approximation. Moreover, define the unit vector $\bp_{B}$ as 
\be \lb{qN}
\bp_{B} = \left(2\sin \frac{\alpha}{2} \right)^{-1} (\bn_B + \bN).
\ee

Taking the limit of Eqs. (\ref{M1}) and (\ref{laJn}), and noting that $ \lim_{r_A \rightarrow \infty}(r_A + r_B - R_{AB}) = r_{c} \tan \alpha/2$ and that $\bN_{AB} = - \bN$, we get
\be \lb{dev}
\underline{\bla} (\bN , \bx_B)  (\gamma+1)\, \frac{2GM}{c^2r_c}\left[\cos \frac{\alpha}{2} \, \bp_{B} +  \sum_{n = 2}^{\infty} J_n \left(\frac{r_e}{r_c}\right)^n  \cos^{n+1} \frac{\alpha}{2}\, {\bm \Lambda}_{n}(\bN , \bx_B)\right],
\ee
where
\bea 
& & {\bm \Lambda}_{2}(\bN , \bx_B) = \left[1 - (\bk . \bn_{B})^{2} - 4 (\bk . \bp_{B})^2\right]\bp_{B} + 2 (\bk . \bp_{B})\left[\bk - (\bk . \bn_{B})\bn_{B} \right] \nonumber \\
& &\qquad \qquad \qquad \, + \sin\frac{\alpha}{2}\left\{ 2 (\bk . \bn_{B})\bk + \left[1 - 3(\bk . \bn_{B})^{2}\right] \bn_{B}\right\}  \lb{lNJ2}
\eea
and
\bea 
& &{\bm \Lambda}_{n} (\bN , \bx_B) = 2^{n-2}(\bk . \bp_{B})^{n-2}\left\{(n-1)[1 - (\bk . \bn_{B})^{2}] - 4 (\bk . \bp_{B})^2\right\}\bp_{B} \nonumber \\
& &\qquad \qquad \qquad   - 2^{n-3}(\bk . \bp_{B})^{n-3} \left\{(n-2)[1 - (\bk . \bn_{B})^{2}] - 4 (\bk . \bp_{B})^2\right\}\left[\bk - (\bk . \bn_{B})\bn_{B} \right] \nonumber \\
& &\qquad \qquad \qquad   + 2^{n-2}\sin\frac{\alpha}{2} (\bk . \bp_{B})^{n-2}\left\{ 2 (\bk . \bn_{B})\bk + \left[1 - 3(\bk . \bn_{B})^{2}\right] \bn_{B}\right\} \nonumber \\
& &+\sum_{m=3}^{n}\, \sideset{}{'}\sum_{i_{1},...,i_{m}}\! \Phi_{nm}(i_1, ..., i_m) \Bigg\lbrack \prod_{l=2}^{m}\left[C_{l}^{(-1/2)}(\bk . \bn_{B}) \right]^{i_{l}}
\left\{(n - m +1)(\bk . \bp_{B})^{i_{1}} \bp_{B} - \frac{i_1}{2} (\bk . \bp_{B})^{i_{1}-1} \left[\bk - (\bk . \bn_{B})\bn_{B} \right]\right\}  \nonumber \\
& & +\sin\frac{\alpha}{2}(\bk . \bp_{B})^{i_{1}}\sum_{l=2}^{m}i_{l}\left[C_{l}^{(-1/2)}(\bk . \bn_{B}) \right]^{i_{l}-1} \prod_{\stackrel{q=2,}{q\neq l}}^{m}\left[C_{q}^{(-1/2)}(\bk . \bn_{B}) \right]^{i_{q}}\left[P_{l-1}(\bk . \bn_{B})\bk - P_{l}(\bk . \bn_{B})\bn_{B}\right]\Bigg\rbrack \Bigg\rbrace \lb{lNJn}
\eea
for $n\geq 3$, the coefficients $\Phi_{nm}(i_1, ..., i_m)$ being defined by
\be \lb{Phip}
\Phi_{nm}(i_1, ..., i_m) = \frac{1}{2}(-1)^{n-m+i_{1}} 2^{m + i_{1}}\frac{(n-m)!}{i_{1}! ... i_{m}!}\sin^{i_{3} + ... +(m-2)i_{m}}\frac{\alpha}{2}.
\ee

The contributions of the multipole moments to the deflection of light will be completely negligible in missions like GAIA or SIM except for light rays passing quite nearly the deflecting body. This implies that one can put $\cos \alpha /2 = 1$ and neglect the terms involving $\sin \alpha /2$ in the range where the multipole expansion yielded by Eqs. (\ref{dev}), (\ref{lNJ2}) and (\ref{lNJn}) is relevant. To justify this approximation, we may content ourselves with discussing Eq. (\ref{lNJ2}) since the quadrupole moment of the giant planets prevails over the higher multipole moments. Considering the case of Jupiter, we have $r_{e} = 7.149 \times 10^{4}$ km and  $J_{2} =  0.014736$ \cite{connaissance}, which implies $4GM/c^2 r_{e} J_{2} = 240$ $\mu$as. As a consequence $4GM/c^2 r_{c} J_{2}(r_e/r_c)^2 > 1$ $\mu$as if and only if (iff) $r_c < 6.2 r_e$. This last inequality implies that the influence of $J_{2}$ may be taken into account iff $\sin \alpha /2 <  5\times 10^{-4}$ since the distance $r_{B}$ between Jupiter and GAIA or SIM is always $ > 6 \times 10^{8}$ km. Then $\vert 1-\cos \alpha/2 \vert <2.5 \times 10^{-7}$ and the contribution of the term containing $\sin \alpha/2$ in the r.h.s. of Eq. (\ref{lNJ2}) is markedly less than one $\mu$as. A similar conclusion can be drawn for $n\geq 3$. As a consequence, in GAIA or SIM missions, the formulae obtained in this section yield results which do not significantly differ from the approximation obtained by assuming that the observer is at infinity, as we shall see below.

\section{Source and observer located at infinity}

Let $P$ be the foot of the perpendicular drawn to ${\cal D}$ from $O$. Since $r_{c} = \vert \bx_{P} \vert$, we may define the unit vector $\bp$ as
\be \lb{bp}
\bp = \frac{\bx_{P}}{r_{c}}.
\ee
It is easily seen that $\bp_{B}$ is given by
\be \lb{pB}
\bp_{B} = \cos \frac{\alpha}{2} \bp + \sin \frac{\alpha}{2} \bN.
\ee

It follows from Eq. (\ref{imp}) that the case where point $B$ is at infinity on $({\cal D})$ corresponds to $\alpha = 0$. So, we have $\lim_{r_{B}\rightarrow \infty}\bp_{B} = \bp$ and we can put
\be \lb{}
\underline{\bla} (\bN, \bp, r_c) = \lim_{r_{B}\rightarrow \infty} \underline{\bla} (\bN , \bx_B).
\ee
The corresponding limit of each term ${\bm \Lambda}_{n}(\bN , \bx_B)$ is obtained by replacing $\alpha$ by $0$, $\bn_{B}$ by $-\bN$ and $\bp_{B}$ by $\bp$ in Eqs. (\ref{lNJ2}) and (\ref{lNJn}). It follows from (\ref{Phip}) that only the solutions to Eqs. (\ref{bruno23}) such that $i_3 + 2 i_4 + ... + (m-2)i_m =0$ have to be retained in the sum $\sideset{}{'}\sum_{i_{1},...,i_{m}}$. A simple calculation leads to the multipole expansion  
\be \lb{lNpc}
\underline{\bla} (\bN , \bp, r_{c})  = (\gamma + 1) \frac{2GM}{c^2 r_c}\left[\bp  + \sum_{n=2}^{\infty} J_{n} \left(\frac{r_e}{r_c}\right)^{n} {\bm \Lambda}_{n} (\bN , \bp, r_{c})\right],
\ee
where
\bea \lb{lNJnb}
& &{\bm \Lambda}_{n} (\bN , \bp, r_c) =  \sum_{m=1}^{p_n}(-1)^{m} \frac{2^{n-2m+1}(n-m)!}{(n-2m+2)!(m-1)!} \left[1-(\bk.\bN)^2\right]^{m-1}\Big\lbrace 2(n-m+1)(\bk.\bp)^{n-2m+2}\bp \nonumber \\
& &\qquad \qquad \qquad \quad  - (n-2m+2)(\bk.\bp)^{n-2m+1}\left[\bk -(\bk.\bN)\bN \right]\Big\rbrace
\eea
for any $n \geq 2$, $p_{n}$ being the integer defined by
\begin{equation} \lb{pn}
p_n=\frac{n}{2}+1\;\mbox{if $n$ is even,}\qquad p_n=\frac{n+1}{2}\;\mbox{if $n$ is odd.}
\end{equation}

In order to discuss the bending of light rays, it is convenient to introduce the orthonormal triad formed by $\bp, \bN$ and 
\be \lb{}
\bq = \bp \times \bN.
\ee
Noting that $\bk - (\bk . \bN)\bN = (\bk . \bp) \bp + (\bk . \bq) \bq$ and that $1-(\bk . \bN)^2 = (\bk . \bp)^2 + (\bk .\bq)^2$, we find that Eq. (\ref{lNJnb}) may be written as
\bea \lb{lNJnbi}
& &{\bm \Lambda}_{n} (\bN , \bp, r_{c}) =  \sum_{m=1}^{p_n}(-1)^{m} \frac{2^{n-2m+1}(n-m)!}{(n-2m+2)!(m-1)!} \left[ (\bk . \bp)^2 + (\bk .\bq)^2\right]^{m-1} \left[ n (\bk.\bp)^{n-2m+2} \bp \right. \nonumber \\
& &\left. \qquad \qquad \qquad \quad - (n-2m+2)(\bk.\bp)^{n-2m+1}(\bk . \bq) \bq \right]. 
\eea

Equation (\ref{lNJnbi}) yields for $n=2, 3, 4$: 
\bea
& &{\bm \Lambda}_{2}(\bN, \bp, r_{c}) = \left[(\bk.\bq)^2 - (\bk.\bp)^2 \right]\bp + 2(\bk.\bp)(\bk.\bq)\bq , \label{J2crosta} \\
& &{\bm \Lambda}_{3}(\bN, \bp, r_{c}) =  (\bk.\bp)\left[ 3 (\bk.\bq)^2 - (\bk.\bp)^2\right] \bp + (\bk.\bq)\left[ 3(\bk.\bp)^2 - (\bk.\bq)^2\right] \bq, \label{J3}\\
& &{\bm \Lambda}_{4}(\bN, \bp, r_{c}) =  \left[6(\bk.\bp)^2(\bk.\bq)^2 - \left(\bk.\bp\right)^4 - (\bk.\bq)^4 \right]\bp + 4(\bk.\bp)(\bk.\bq)
\left[(\bk.\bp)^2 -(\bk.\bq)^2\right]\bq \label{J4}.
\eea

Suppose that the observer is at rest at space infinity. This observer sees the light source in the direction $\bN + \bDa$ determined by the opposite of the space-like contravariant components of the vector tangent to the light ray. It follows from Eq. (\ref{tan}) and $\bN_{AB} = -\bN$ that  $\bDa = \underline{\bla} (\bN, \bp, r_c)$. So the effectively observed deflection vector is given by the multipole expansion
\be \lb{bDa}
\bDa = \bDa_{M}(\bN , \bp, r_{c}) + \sum_{n=2}^{\infty}\bDa_{J_{n}}(\bN , \bp, r_{c}),
\ee
where 
\bea 
\bDa_{M}(\bN , \bp, r_{c}) &=& (\gamma + 1) \frac{2GM}{c^2 r_c} \bp, \lb{bDM} \\
\bDa_{J_{n}}(\bN , \bp, r_{c}) &=& (\gamma + 1) \frac{2GM}{c^2 r_c} J_{n} \left(\frac{r_e}{r_c}\right)^{n} {\bm \Lambda}_{n} (\bN , \bp, r_{c}). \lb{bDJn}
\eea

It follows from Eqs. (\ref{bDM})-(\ref{bDJn}) and (\ref{lNJnbi}) that $\bDa$ is orthogonal to $\bN$.

Substituting for ${\bm \Lambda}_{2}$ from Eq. (\ref{J2crosta}) into Eq. (\ref{bDJn}) yields the expression of the deflection vector $\bDa_{J_{2}}$ obtained in previous works (see, e.g., Refs. \cite{2006CQGra..23.4853C} and \cite{2007PhRvD..75f2002K}).

If line ${\cal D}$ lies in the equatorial plane, then $\bk . \bp = 0$ and $\bq = \pm \bk$. As a consequence Eq. (\ref{lNJnbi}) gives
\be \lb{Jle}
\bDa_{J_{2k}}(\bN , \bp, r_{c}) = (-1)^{k+1}(\gamma + 1)\frac{2GM}{c^2 r_{c}} J_{2k} \left(\frac{r_{e}}{r_{c}}\right)^{2k} \bp
\ee
for the mass multipole of even order $2k$ and 
\be \lb{Jlo}
\bDa_{J_{2l+1}}(\bN , \bp, r_{c}) = (-1)^{l}(\gamma + 1)\frac{2GM}{c^2 r_{c}} J_{2l+1} \left(\frac{r_{e}}{r_{c}}\right)^{2l+1} \bk
\ee
for the mass multipole of odd order $2l + 1$. We note that $\bDa_{J_{2k}}$ and $\bDa_{J_{2l+1}}$ are orthogonal whatever $k$ and $l$. 

If ${\cal D}$ is in a meridian plane, then $\bk.\bq = 0$. Pointing out that the relation
$$ 
\sum_{m=1}^{p_n}(-1)^{m} \frac{2^{n-2m+1}(n-m)!}{(n-2m+2)!(m-1)!}\, n = -1
$$
is valid whatever $n \geq 1$ \footnote{This relation can be checked with MAPLE 11.}, it may be seen that Eq. (\ref{lNJnbi}) reduces to
\be \lb{Jnm}
\bDa_{J_{n}}(\bN , \bp, r_{c}) = - (\gamma + 1) \frac{2 GM}{c^2 r_c} J_{n} \left(\frac{r_e}{r_c}\right)^{n} (\bk . \bp)^n \bp.
\ee
 
It follows from Eq. (\ref{Jnm}) that for a ray propagating in a meridian plane the greatest deflecting effect due to the $J_{n}$ occurs when $\bp = \pm \bk$, i.e. when ${\cal D}$ is parallel to the equatorial plane. Equation (\ref{Jnm}) shows also that the mass multipole moments have no deflection effect when the direction of emission at infinity is parallel to the axis of symmetry, since $\bk . \bp =0$ in this case. 

The deflection angle $\widehat{\delta}(\bN , \bp, r_{c})$ is defined as the angle between $\bN$ and $\bN + \bDa(\bN , \bp, r_{c})$.  Since $\bN$ is a unit vector and $\bN . \bDa = 0$, we have
\be \lb{dev1}
\vert \widehat{\delta}(\bN , \bp, r_{c}) \vert = \vert \bDa(\bN , \bp, r_{c})\vert + \mathcal{O}\left(\frac{1}{c^4}\right).
\ee

Let us briefly examine the contributions of $\widehat{\delta}(\bN , \bp, r_{c})$ due to $J_{2}$, $J_{3}$ and $J_{4}$. It follows from Eqs. (\ref{J2crosta})-(\ref{J4}) that  
\be \lb{dev3}
\vert \bDa_{J_{n}}(\bN , \bp, r_{c}) \vert = (\gamma + 1) \frac{2 GM}{c^2 r_c} J_{n} \left(\frac{r_e}{r_c}\right)^{n} [1 - (\bk . \bN)^2]^{n/2}
\ee
for $n= 2, 3, 4$ \footnote{It may be conjectured that Eq. (\ref{dev3}) is valid for any $n$.}. In each of these cases, the highest possible value for $\vert \bDa_{J_{n}}(\bN , \bp, r_{c}) \vert$  is reached when $\bN$ is orthogonal to $\bk$. As a consequence we have 
\be \lb{devm}
\vert \bDa_{J_{n}}(\bN , \bp, r_{c}) \vert_{\mbox{max}} = (\gamma + 1) \frac{2 GM}{c^2 r_c} J_{n} \left(\frac{r_e}{r_c}\right)^{n}.
\ee

For Jupiter, $J_2 = 0.014736$ (see Sec. VI), $J_3= 0.000001$ and $J_4= - 0.000587$ \cite{connaissance}. With $\gamma =1$, the predicted deflexions of a grazing ray specifically due to $J_2$, $J_3$ and $J_4$ are then in the range
\be \lb{Jup}
\vert \widehat{\delta}_{J_{2}} \vert \leq 240 \, \mu\mbox{as}, \quad \vert \widehat{\delta}_{J_{3}} \vert \leq 0.016 \, \mu\mbox{as}, \quad \vert \widehat{\delta}_{J_{4}} \vert \leq  9.6 \, \mu\mbox{as}.
\ee

So our formulae yield a rigorous confirmation of previous estimates given in Ref. \cite{Klioner:2003a}.  The lowest possible value for $\vert \bDa_{J_{n}}(\bN , \bp, r_{c}) \vert$ is $0$ and corresponds to the case where $\bN = \pm \bk$.      

\section{Conclusion}
\par This work yields a complete determination at the order $1/c^2$ of the bending of light in a static gravitational field generated by an isolated axisymmetric body. The method developed here could be extended to the contributions of the spin multipole moments. The explicit formula (\ref{TJn}) giving the multipole expansion of the time transfer function may also be of interest for the analysis of the frequency transfers between two atomic clocks. 


\end{document}